\documentclass[sigconf]{acmart}

\copyrightyear{2026}
\acmYear{2026}
\setcopyright{cc}
\setcctype{by}
\acmConference[IDE '26]{3rd International Workshop on Integrated Development Environments }{April 12--18, 2026}{Rio de Janeiro, Brazil}
\acmBooktitle{3rd International Workshop on Integrated Development Environments (IDE '26), April 12--18, 2026, Rio de Janeiro, Brazil}
\acmPrice{}
\acmDOI{10.1145/3786151.3788600}
\acmISBN{979-8-4007-2384-1/2026/04}

\usepackage{listings}
\usepackage{url}

\begin{document}

\title{Optimizing an IDE for an Evolving Language Ecosystem}

\author{Adam Welc}
\affiliation{
  \institution{Mysten Labs}
  \city{Palo Alto}
  \country{USA}
}
\author{Todd Nowacki}
\affiliation{
  \institution{Mysten Labs}
  \city{Palo Alto}
  \country{USA}
}
\author{Dario Russi}
\affiliation{
  \institution{Mysten Labs}
  \city{Palo Alto}
  \country{USA}
}
\author{Cameron Swords}
\affiliation{
  \institution{Mysten Labs}
  \city{Palo Alto}
  \country{USA}
}
\author{Timothy A. K. Zakian}
\affiliation{
  \institution{Mysten Labs}
  \city{Palo Alto}
  \country{USA}
}

\renewcommand{\shortauthors}{Adam Welc, et al.}

\ccsdesc[500]{Software and its engineering~Software maintenance tools}
\ccsdesc[500]{Software and its engineering~Compilers}

\keywords{programming language, compiler, IDE, optimization}

\begin{abstract}
This paper describes a strategy for developing a high performance and feature-rich IDE for an
evolving smart contract language ecosystem. Our target is Move, a programming language for the Sui
smart contracts platform. The strategy we chose to support the Move language ecosystem utilizes
Language Server Protocol (LSP) and it is based on the already existing ``core'' language
machinery, in particular the core language compiler. We discuss alternatives we considered, as well
as the evolution of our infrastructure that was necessary to keep up with the growth of the
language ecosystem, particularly with respect to optimizations (and their impact) that needed to be
implemented to accommodate this growth. We conclude with lessons learned during the IDE support
development process that we hope will be beneficial for others attempting to follow a similar path.
\end{abstract}

\maketitle

\section{Introduction}

Creating a new programming language is hard. Driving adoption of a new language may be even
harder. While programming language features such as expressibility, security and performance are
essential to generate interest in a new language, it is nearly impossible to achieve wide adoption
without proper tooling. In particular, providing access to a feature-rich Integrated Development
Environment (IDE) results in a lower barrier to entry for new developers and reduces friction when
writing code using potentially unfamiliar new constructs or libraries. IDE features such as code
navigation capabilities (e.g., finding a type or declaration of a given language construct), code
comprehension (e.g., additional information on language constructs provided when hovering over
them), smart code auto-completion, syntax highlighting, or visualization of compiler diagnostics in
the code editor are particularly helpful.

Ideally, when deciding to create a new programming language, you assemble a large team to tackle
both language design and implementation but also all relevant developer-facing tooling including an
IDE. Then you release both the ``core'' language machinery (e.g., a compiler and a virtual machine) and
mature developer-facing tooling at the same time. Practically, resources are typically limited, and
core language machinery often takes precedence, as it not only has to be of sufficiently high
quality at the point of the language's initial release, but also needs to be maintained and extended
as the language inevitably evolves and its developer community grows.

This is the situation we found ourselves in at Mysten Labs on a team responsible for maintaining and
extending Move\cite{move}, a programming language for the Sui\cite{sui} smart contracts
platform. The language itself and its supporting infrastructure have been designed and implemented
from scratch to provide state-of-the-art for smart contract programming in terms of expressiveness
and security. Move was originally created at Facebook, so our starting point was having an initial
version of the compiler and virtual machine, as well as a skeleton IDE implementation whose most
useful feature was syntax highlighting.

This paper focuses on our quest to deliver a feature-rich and high performance IDE to our
developers. While this journey is in some ways specific to our particular set of constraints, as
well as our language and the implementation of its core machinery, we believe that experience we
gathered and lessons we learned in the process are general enough and interesting enough to be
useful for the general IDE research and development community.
This was in fact one of the main motivations for writing this paper as we found it really difficult
to find any consolidated information on how one should proceed when developing IDE support for an
evolving language ecosystem, and what the tradeoffs between choosing different strategies could
be. This is also the reason why this paper does not have a formal section describing related work,
as all we had at our disposal was fragmented information and source code examples.

In this paper we make the following contributions:
\begin{itemize}
  \item We describe our initial development strategy, alternatives we considered, and what were
    the main factors driving our particular choice of the development strategy.
  \item We explain how IDE support needed to evolve, particularly in terms of a series of
  optimizations that were required to keep up with the growth of the Move language ecosystem. We
  describe optimizations we implemented and their impact on IDE performance.
\end{itemize}

All code referred to in this paper is open source and available at
\url{https://github.com/MystenLabs/sui}.

\section{IDE development strategy}

Programmers have multiple IDEs at their disposal. Ideally, and with unconstrained resources, one
would like to support all of them right from the start. In practice, one has to decide how to
maximize impact while minimizing development effort. The informal metric we chose to drive this
decision was a desire to reach the highest number of programmers by targeting IDEs of their
choice. Cross-platform IDEs provided by JetBrains (a family of IDEs\cite{jetbrains} both free and
commercial, powered by the same engine) and by Microsoft (Visual Studio Code\cite{vscode} and Visual
Studio\cite{vstudio}) are generally considered\cite{idesurvey,popularides,topides} to be the most
popular ones, with others (e.g, Eclipse\cite{eclipse}) still widely used but not reaching
quite the same level of popularity. It would then be logical to target either JetBrains or Microsoft
products, with the choice being potentially difficult, except for one additional
consideration.

Language support for JetBrains products is typically provided natively using APIs provided by
JetBrains, while Microsoft products utilize an open solution, Language Server Protocol (LSP)\cite{lsp},
to support virtually all advanced IDE features. This protocol is supported by many of the
third-party IDEs and editors (e.g., Eclipse, Sublime\cite{sublime}) which can greatly extend developer
reach. Even JetBrains products include support for the LSP but native support for this protocol is
as of this point restricted to commercial versions of their product line. This made our
choice of high-level implementation strategy simple. We settled on an LSP-based
implementation and decided to initially target Visual Studio Code (VSCode), which is available for
free, as our main integration platform (our developer community additionally supports integrations
with Emacs\cite{emacs}, Vim\cite{vim}, and Zed\cite{zed}). LSP keeps evolving and its current set of
supported features has largely satisfied our requirements so far. In the future, we may consider
enhancements that go beyond the LSP feature set even if it improves user experience only for some
integrations. For example, some nice-to-haves for the VSCode integration would be to show analysis
progress in the status bar or show full compiler-formatted error and warnings in separate tabs.

The second choice we had to make was on the lower-level implementation strategy. The LSP
implementation consists of two components, an LSP client (provided by the IDEs supporting LSP)
and an LSP server (needs to be implemented by the creators of the new language). An LSP
client is responsible for requesting information about user code from the LSP server, so that it can
be presented to the programmer by the IDE's user interface. For example, with the IDE's code editor
cursor on some language construct, the LSP client can ask the LSP server for specific information about
this construct on behalf of the user -- with a cursor on a variable name, LSP client can ask for
this variable's type and display it in a pop-up window, or ask where the definition of this variable
is located to jump directly to this location. An LSP server, in order to provide this kind of
information, must have a deep semantic understanding of the user code and its structure.

In an ideal world one would build a dedicated analyzer from scratch that is best suited for user
code analysis in a context of an LSP server. This approach is certainly feasible given sufficient
resources, and has proven to be quite successful, with one of the prime examples being
rust-analyzer\cite{rustanalyzer}. On the other hand, unless continued support and maintenance are somehow
guaranteed, such dedicated solution may easily fall behind evolution of the language and its
community growth.

An alternative, one that we ultimately decided to follow, was to utilize as much of the existing
core language machinery as possible, in particular to use the language's compiler. While a core
language compiler, whose main purpose is to generate executable code, is unlikely to provide all the
information that an LSP server implementation needs to serve to the client, it certainly has a deep
understanding of the code it must compile (e.g., functions, types, variables). It should then form a
reasonable LSP server implementation starting point. Additionally, our initial performance
experiments indicated sub-second compilation times which was deemed sufficiently fast to provide good
user experience in the context of and IDE.

In Section~\ref{sec:evolution} we describe our initial implementation of the LSP server and discuss
how it needed to evolve to keep up with the language ecosystem growth, particularly to handle larger
and more complicated code bases. In Section~\ref{sec:perf} we present empirical evaluation results
explaining evolution of our implementation with respect to its performance in more detail.

\section{IDE support evolution}
\label{sec:evolution}

In this section we describe an initial implementation of our LSP server based on the core Move
language compiler and discuss how this implementation choice drove infrastructure changes to stay
abreast of the language ecosystem evolution. We intentionally keep the implementation description at
a high level as the low-level details are specific to our internal code base, and are unlikely to
be of much interest to the broader community.

\subsection{Initial implementation}

The Move language flavor being developed at Mysten Labs is an evolution of its original form created
at Facebook. It uses objects to represent user assets which is a natural match for the data model of
the Sui smart contract platform. Its syntax resembles Rust, and one of its main safety guarantees is
inspired by Rust's data ownership to prevent accidental object copying or discarding. The code is
organized into modules that are then organized into packages, promoting modularity and code reuse.

The core Move
compiler~\footnote{\url{https://github.com/MystenLabs/sui/tree/main/external-crates/move/crates/move-compiler}}
is a multi-pass optimizing compiler translating Move source code to Move
bytecode (which is then executed by the Move virtual machine). At each compilation pass, the
compiler produces intermediate representation of the program in a form of an Abstract Syntax Tree
(AST). The further from the first pass, that is parsing of the source code into the first level of
an AST, the more abstract the ASTs become, which can lead to loss of some information, such as
source code structure. On the flip side, ASTs generated later during compilation include previously
unavailable information, such as information about types. As a result, our
implementation of the LSP server taps into compiler ASTs at parsing level and at typing level. Most
of the required information is available in the typing-level AST, for example types or variable and
function uses and definitions. Locations are preserved throughout different AST levels which
simplifies collection or relevant information. We need parsing-level AST only to collect information about language
constructs that get abstracted away by subsequent compilation passes. For example, function and type
imports from different modules are no longer present at the typing-level AST but analysis needs to
understand their structure for code navigation (to support navigating to a module where they are
defined) and for correct smart auto-completion.

The Move compiler is written in Rust as is our LSP server
implementation. During initial tests on some of our internal code, compilation portion of serving a single
LSP request (such as finding a definition of a variable) was on the order of 400ms, and it was the
most expensive part, with the analysis of the ASTs being an order of magnitude less expensive (around 40ms).
We deemed this to be an acceptable performance level to deliver good user experience in terms of IDE's response.
Conversely, anything above 1s mark would be unacceptable as it would lead to a much more noticeable delay.
However, with the Move language ecosystem growth, the size of the code that needed to be
processed inside an IDE grew as well, and compilation times of the initial implementations started
to slip, requiring introduction of additional compilation time optimizations.

\subsection{Compilation time optimizations}
\label{sec:comp}

Initial release of the Move language came with a rather rudimentary set of standard libraries, but
as the ecosystem evolved, so did the libraries. In our original implementation, due to how the Move
compiler was architected, both user code and its dependencies (i.e., libraries) were compiled each
time the IDE requested updated information (e.g., after a code modification). With the size of the
dependencies growing, we started observing compilation times approaching 1s and realized that
situation will only get worse over time.

Fortunately, in the IDE setting, only the user code gets modified during program development, and
there is no need to compile (and analyze) dependencies each time (only) the user code
changes~\footnote{If user explicitly changes the dependencies, this has to be detected, and the
dependencies must be re-compiled and re-analyzed.}. Even more fortunately, the Move language
compiler already included a facility to pre-compile dependencies and re-use them during subsequent
compilations, otherwise we would be forced to implement this type of machinery from scratch. It was
originally implemented to improve performance of compiler unit tests but was easy to adapt to work
in the context of the LSP server implementation, and brought the overall compilation time down to
significantly less than 1ms.

As the number of programmers for a given language grows and the language ecosystem evolves, the size
of user code grows as well. Even with pre-compiled dependencies, over time we observed compilation times
growing back up closer to 1s range. The saving grace here is that only a small fraction of user
code gets modified at any given time during a programming session. Conceptually, we should be able to use a
similar mechanism to the one applied when handling libraries, but the compiler did not support it at the right
granularity level to handle individual files from the user code. Instead, we modified the compiler
to only ``fully'' compile modified user files and ``partially'' compile unmodified user files, where
partial compilation would skip compiling function bodies. This way the compiler would still have a
view of the entire user program but overall compilation time would be significantly reduced. In
order for this to work, we needed to also cache analysis results for partially compiled files (as
otherwise we would have no function bodies to analyze) while performing ``regular'' analysis for
fully compiled files. This optimization brought overall compilation time back to less than 200ms. 

Compilation time directly affects experience of IDE users and optimizing it was our main priority
even at the cost of increased memory consumption, but eventually optimizing memory utilization became
a necessity as well.

\subsection{Memory utilization optimizations}

The library code caching helped significantly reduce overall compilation times, as discussed in
Section~\ref{sec:comp}, but the machinery we used to implement this was not optimized towards memory
consumption as it was originally only used during testing. In short, the data structure to store
pre-compiled dependencies contained a lot of information that was not strictly necessary to enable
compilation against these dependencies -- it included ASTs for all compiler passes as well as the
final compiled bytecode. As a result, LSP server's memory footprint for a single Move
package~\footnote{Move code is organized into modules that are then organized into packages.} could
reach more than 1.5GB for larger user packages (and more than 1.3GB for smaller ones).
With programmers opening multiple packages in a single workspace, the memory overhead could become
truly overwhelming. The solution was to re-architect portion of the compiler responsible for
handling pre-compiled libraries to only include information absolutely necessary for compilation
(data types, function signatures, bodies of selected functions only needed for macro inlining,
etc.). This allowed us to reduce memory footprint for a single package to a little over 800MB for
larger packages (and to a little over 650MB for smaller ones). There is likely some remaining room
for improvement there but we considered an almost 100\% reduction a win.

The memory footprint reduction resulting from re-architecting compiler-level data structure
holding pre-compiled dependencies was quite successful but did not address an issue of programmers
opening multiple packages inside the same workspace, some of which may share the same set of
dependencies (e.g., standard library). This is because caching so far has been done on a per-package
basis. The final optimization was then to re-architect the cache structure to share dependencies
among multiple packages. We implemented this at the dependent package granularity level, so that
even if user packages cannot share all dependencies, they can at least share some (e.g., standard
library packages). This optimization provided a more modest improvement -- according to our
measurements for 3 packages sharing our standard library code, we obtained around 15\% reduction in
memory footprint when all these packages were opened inside the same workspace.

\section{Performance}
\label{sec:perf}

In Section~\ref{sec:evolution} we discussed compiler optimizations needed to keep up with an
evolving language ecosystem along with some performance numbers indicating their impact. In
this section we present a more detailed description of the impact of various optimizations based on
an empirical evaluation of the LSP server's performance at various stages of its development.

\subsection{Methodology}

We evaluate LSP server performance at 4 pivotal points during its development when different
optimizations were introduced. Along with a short description, we include git commit SHA in our
repository (\url{https://github.com/MystenLabs/sui}) reflecting the point in time when each
optimization was introduced.
\begin{itemize}
\item \textbf{pre-compiled} -- pre-compiled dependencies\\(SHA:
  \texttt{f0240a534ff0512975b8d3358770141d55cf4a4c})
\item \textbf{incremental} --``incremental'' compilation of user code via skipping compilation of
  unmodified function bodies\\(SHA: \texttt{ac0361801e151ff5007aef033702b10438755ba1})
\item \textbf{lean-deps} -- a leaner version of pre-compiled dependencies\\(SHA:
  \texttt{ac0361801e151ff5007aef033702b10438755ba1})
\item \textbf{cross-package} -- caching dependencies across packages\\(SHA:
  \texttt{acb70d3b83695da14abbc3f0fe5ffb3be2e9b8a4})
\end{itemize}

To collect the actual numbers with minimal interference from an IDE we built a testing
harness consisting of an LSP client responsible for communicating with the LSP server, light
instrumentation in the LSP server to collect compilation times, and a driver script which was also
responsible for collecting memory consumption information utilizing the \texttt{top} command.  All
measurements were collected on a MacBook Pro with the Apple M4 Max chip and 64GB or RAM running
macOS Sequoia v15.6.1.

\subsection{Evaluation}

We collect all data for each optimization for a version of the LSP server right before a given
optimization was in place (i.e.., git commit SHA prior to the optimization being implemented) and
right after it was in place (i.e., git SHA of the commit implementing the optimization).

When evaluating a single package, we run 20 ``warmup'' iterations followed by 20 ``measurement''
iterations, and report an average over the latter. Each iteration is triggered by a ``file modified''
event sent to the LSP server by the testing harness. When evaluating multiple packages in the same
workspace, and we do it only for \textbf{cross-package} optimization to collect memory footprint
data, we run 20 warmup and 20 measurement iterations for each package in turn but report average
memory footprint only for the last 20 measurement iterations for the last package being evaluated
(so that cumulative numbers for all packages are included).

It is important to understand that the comparison between different sets of numbers only make sense
pair-wise. In other words, it is meaningful to compare results between the time right before a given
optimization was implemented and the time right after this optimization was implemented, but not
to compare numbers between different optimizations. There are several reasons for it. One of them is
that we cannot always use the same user code for evaluation (as earlier version for of the LSP server
do not support newer and larger code bases). Another one is that the LSP server was in active development
between the times when these optimizations were implemented, with new features added, bugs fixed,
etc., that could easily affect overall performance.

We used two different applications for evaluation, both developed internally and both available in
open source:
\begin{itemize}
\item \textbf{capy} (\url{https://github.com/MystenLabs/sui}) -- a marketplace for
  capybara NFTs that used to be a mascot for our platform; this application is supported by all
  versions of the LSP server and represents the kinds of applications that were available at the
  earlier stages of the Move language ecosystem evolution (we used a version at git commit SHA
  \texttt{020ef08424309680fd5b73626b8f3ecad136b27d} with total 1066 of lines of code, excluding
  empty lines)
\item \textbf{deepbook} (\url{https://github.com/MystenLabs/deepbookv3}) -- a
  decentralized central limit order book (at git
  commit SHA \texttt{9ed8ab323451626694f9ea7ebddafbb66bbf61d3}); it represents a much newer,
  larger and more complicated code base (total number of lines of code, including tests, and
  excluding empty lines, is 19433)
\end{itemize}

The numbers for different optimizations are presented below.

\paragraph{\textbf{pre-compiled}}

This optimization was implemented early on and for its evaluation we use the \textbf{capy}
application. When including all standard library dependencies available in our repository at the
time of this LSP server optimization being implemented, compilation time for the \textbf{capy} was
\textbf{0.924s} prior to the optimization being implemented and \textbf{0.076s} after it was
implemented.

\paragraph{\textbf{incremental}}

At this stage of the LSP server development we could use larger and newer code base and this
optimization was evaluated based on compilation performance of the \textbf{deepbook} application:
prior to the optimization being implemented it was \textbf{0.703s} and after this optimization was
in place it was \textbf{0.164s}.

\paragraph{\textbf{lean-deps}}

At the time this optimization was being considered, our build system was re-architected to include
implicit standard dependencies in every build. This significantly increased memory pressure for all
applications, regardless of the size. In particular, prior to this optimization being implemented,
LSP memory footprint for the \textbf{capy} application was \textbf{1.34GB} and for the
\textbf{deepbook} application was \textbf{1.47GB}. After this optimization was implemented, it
dropped down to \textbf{664.0MB} and \textbf{826.3MB}, respectively.

\paragraph{\textbf{cross-package}}

To evaluate this last optimization, we combine the \textbf{capy} and \textbf{deepbook} applications
into a single workspace. The test harness triggers three sets of 40 iterations (20 warmup ones and
20 measurement ones) to three different packages in the same workspace in the following order:
\textbf{capy} package, \textbf{deepbook} package, and \textbf{token} package which is an internal
dependency of the \textbf{deepbook} package. Measurements for the \textbf{token} package are
collected last and represent cumulative memory consumption for all three packages. Prior to this
optimization being implemented, cumulative memory footprint for all packages was \textbf{2.07GB} and
after this optimization was implemented it got reduced to \textbf{1.75GB}.

\section{Future Work}

While our IDE support includes the most important features, has been extensively optimized, and is widely used,
particularly for a language as young as Move (over 20k installs from VSCode
Marketplace\footnote{\url{https://marketplace.visualstudio.com/items?itemName=mysten.move}}), we
continue working on both performance and new features. More concretely, we are currently
investigating performance discrepancies among the three platforms we support (i.e., macOS, Linux and
Windows) and continue investigating how to make the compiler even more ``lean'' to further reduce memory
footprint. There is also a long tail of IDE features that we need to support to reach
parity with LSPs for more mature languages that have been in development for much longer, such as
rust-analyzer~\cite{rustanalyzer}. These missing features include code refactorings, additional
auto-fixes or support for LSP's CodeLens, as well as proper support for macros.

This last problem is our current priority as while code navigation and code comprehension work for
macros, auto-completion currently does not. Macros are a notoriously difficult problem to deal with
when developing IDE support, with rust-analyzer development team calling them ``disproportionally
hard to support in an IDE''~\cite{macros}. In our case, the main difficulty is that at the point
when we do most of our analysis, that is at the level of typing AST, macros are already expanded and
their original declarations discarded. To properly support macros, we would have to carry
their declarations through compilation passes to make them available at the typing-level AST, which
could have significant impact on memory utilization.

In the future we also plan to explore benefits that artificial intelligence, particularly large
language models (LLMs), can bring to the developer experience in the IDE. We still feel like
providing a solid IDE experience built using more traditional techniques is important to serve as
the source of truth. LLMs provide exciting opportunities, but can also be subject to hallucinations. This
is particularly relevant when providing Move language support, as the language is still young, which
results in limited amount of training data being available, and syntactically similar to Rust, which could lead to
increased number of hallucinations.

Finally, particularly if enough resources become available, we may consider a complete rewrite of
the analyzer, much like what happened with IDE support for Rust, which moved from a solution based
on their core language compiler to a solution based on a separate analysis
infrastructure~\cite{rlsdeprecation}. However, we feel like we have not yet exhausted opportunities related to
tailoring Move's core compiler to the needs of the IDE development. Also, while incremental changes
inevitably introduce at least some amount of technical debt, some of the changes we made benefited
both IDE and core compilation infrastructure, in particular compiler-side memory optimizations.

\section{Conclusions and lessons learned}

In this paper we discussed a strategy for implementing a high performance and feature-rich LSP
support for an evolving Move language ecosystem. We chose to utilize as much of already
existing core language machinery as possible, particularly the core language compiler. This choice
allowed us to release our IDE quickly and evolve performance of the underlying infrastructure over
time as needed. The main lesson learned from this experience is that one does not necessarily have to
implement all the performance optimizations at once, but that the core language machinery should
allow for incremental improvements which may not always be possible due to how it had been
originally architected. In our case, we were lucky to have at least initial versions of the
mechanisms we ultimately needed, but having this knowledge up front may be useful when architecting
one's own LSP support. In particular:
\begin{itemize}
\item one should anticipate a need to cache dependencies of the user code (e.g., the compiler should
  have the ability to pre-compile and store them ``on the side'')
\item one may have to consider incremental compilation (many compilers, especially in their initial
  versions, are whole-program compilers)
\item one has to be wary of the size of data structures used to cache computation results (memory is
  considered cheap these days but it is not infinite)
\end{itemize}
In some settings, one could also perhaps consider a complementary solution to improving performance,
particularly for developers working on less powerful hardware, that is is to run the analyzer in the
cloud on a top-tier server machine and deliver analysis results to the IDE over the network. Besides
obvious trade-offs, such as reliance on low-latency network connection to deliver good developer
experience, these types of solutions work best in a centralized setting, for example when they are
offered by a tech company to its internal developers. In a distributed settings, with many
independent developers working on separate smart contracts, it is difficult for a single entity to
justify the cost of running this kind of service.

Another lesson is that the language compiler used as a foundation for IDE support may have to evolve
beyond implementing additional performance optimizations. For example, as the set of features we
wanted to support in the IDE grew, we had to modify the parser used by the compiler to be more
resilient~\cite{resilience}. In the original Move compiler, upon encountering a parsing error, the
compiler would simply quit, which was a totally acceptable behavior when it was used on the command
line to produce the Move bytecode. On the other hand, in the IDE setting this is far from ideal,
particularly if an error is encountered early in the compilation process, as the code following the
error remains unanalyzed and some IDE features (e.g, code navigation or code comprehension)
will not work there. It would also make it impossible to implement smart auto-completion which was
one of the most requested features. It is because auto-completion organically relies on the
analyzer's ability to ``understand'' incomplete (and thus temporarily incorrect) code to determine
how to complete only partially fleshed out language constructs (e.g., an already written variable
name representing some structure followed by a dot, but without the field name that is subject to
auto-completion or a semicolon to end the line). As a result, we had to partially rewrite the parser
to handle the errors more gracefully -- allow it to continue parsing beyond the point of error and
recover meaningful partial results that could then be handed over to subsequent compiler passes to
continue the analysis.

We hope that our journey to implement LSP support for the Move language is going to be beneficial
for others attempting a similar feat for their own up-and-coming programming language ecosystems.
While problems and solutions discussed here are for obvious reasons somewhat specific to our initial
setting (e.g., initial compiler performance and feature set) we believe that lessons learned from
choosing the initial strategy that we described in this section as well as the compiler
optimizations, whose description we intentionally tried to keep at a relatively high level, could be
applicable to developing IDE support for a different language ecosystem.

All code implementing IDE support for the Move language (both LSP server and VSCode extension
implementations) is available at \url{https://github.com/MystenLabs/sui}.

\bibliographystyle{ACM-Reference-Format}
\bibliography{bibtex}

\end{document}